\documentclass[12pt]{article}
\usepackage{epsfig}
\usepackage{latexsym}
\parskip=\itemsep               
\newcommand{\lwig}{\mbox{\,\raisebox{.3ex}
{$<$}$\!\!\!\!\!$\raisebox{-.9ex}{$\sim$}\,}}

\begin{document}
\title{\Large \bf Infrared Fixed Points and Fixed Lines for Couplings in the Chiral
Lagrangian\\[12mm]}
\author{Mario Atance\,\thanks{Work supported by the spanish FPU programme
under grant PF 97 25438231. E-mail atance@theo-physik.uni-kiel.de}\,
$^{(1)}$\ \ 
and\ \ 
Barbara Schrempp\,\thanks{E-mail schrempp@physik.uni-kiel.de}\,
$^{(2,1)}$ \\[7mm]
$^{(1)}${\small \it Institut f{\"u}r Theoretische Physik und Astrophysik, 
Universit{\"a}t Kiel, D-24118 Kiel, Germany} \\
$^{(2)}${\small \it Institut f{\"u}r Experimentelle und Angewandte Physik, 
Universit{\"a}t Kiel, D-24118 Kiel, Germany}}
\maketitle
\begin{abstract}
\noindent
In the framework of the low energy chiral Lagrangian renormalization
group equations for the couplings are investigated up to order $p^6$ --
as well for $SU(2)\times SU(2)$ as for $SU(3)\times SU(3)$ chiral symmetry.
Infrared attractive fixed points for ratios of $O(p^4)$ couplings are
found, which turn out to agree with the values determined from 
experiment in a surprisingly large number of cases. Infrared
attractive fixed line solutions for $O(p^6)$ couplings in terms
of $O(p^4)$ couplings and among $O(p^6)$ couplings are determined.\\[2mm]
\noindent
PACS number(s): 11.10.Hi, 12.39.Fe
\noindent
Keywords: Chiral Lagrangian, Infrared Fixed Points
\end{abstract}
\bigskip
As is well known, an appropriate extension of QCD to low
energies is the chiral Lagrangian, which realizes the spontaneously 
broken (approximate) chiral symmetry nonlinearly in terms of the light
Goldstone field degrees of freedom\,\cite{wei}. As in
any effective field theory, the Lagrangian has infinitely many 
contributing operators, which may be arranged according to their
importance for low energy observables in an expansion in powers of 
$p/\Lambda$. Here $p$ denotes the low momentum scale of interest and
$\Lambda$ some momentum cut-off, above which the chiral Lagrangian
ceases to be valid, with $p/\Lambda \lwig O(1)$. The number of operators
contributing to each order in this expansion is finite. 
Similarly, perturbation theory in the number of loops and the renormalization
program can be carried out for effective field theories --- even though in 
principle infinitely many counterterms are required. The choice of a 
mass-independent renormalization scheme (MS, $\overline{{\mathrm MS}}$) 
leads, however, to counterterms which may again be arranged in an expansion
in powers of $p/\Lambda$. As a consequence, in any given order in $p/\Lambda$
the number of counterterms needed to absorb the divergences is again 
finite\,\cite{chi}.

Of interest are the coefficients of the operators in the chiral Lagrangian,
or rather --- after extraction of their dimension in form of powers of 
$\Lambda$ --- the dimensionless couplings. These couplings encode in 
principle the information about the QCD dynamics at higher scales.
Unfortunately, in practice they are unknown. There have been efforts to 
estimate some of them by using different techniques (lattice 
calculations, large $N_c$ limit, vector meson dominance, \ldots) \cite{chi}.
The couplings in a given order of $p/\Lambda$ may be determined by
experiment; one needs as many observables as couplings to be determined.
Within the framework of perturbation theory and renormalization described
above, the renormalization group equations for the couplings can be 
determined at any fixed order in $p/\Lambda$. They have all been calculated
up to $O((p/\Lambda)^4)$ -- abbreviated by $O(p^4)$ -- and
recently up to $O(p^6)$\,\cite{nueva} (for earlier work we
refer to references quoted in Refs.\,\cite{nueva},\,\cite{p6}).

In this paper we investigate the renormalization group equations
(RGE) up to $O(p^6)$ and 
search for infrared (IR) attractive fixed point and fixed line solutions. This is a perfectly
legitimate search, since the IR limit probes small momenta, where
the chiral Lagrangian is applicable. The analysis is performed as well for
$SU(2)\times SU(2)$ as for $SU(3)\times SU(3)$ chiral symmetry. As it turns
out, the RGE indeed exhibit IR fixed points
in {\it ratios\/} of $O(p^4)$ couplings as well as IR fixed lines and
surfaces  relating
$O(p^6)$ to $O(p^4)$ couplings and $O(p^6)$ couplings among
each other. These fixed points and higher fixed manifolds are non-trivial special
solutions of the renormalization group equations which do not depend
on any initial values for the couplings and which in addition attract the 
renormalization group flow in its evolution from the scale $\Lambda$ towards
the IR. It is very interesting to see, how they compare to  
experimental data as far as
these are available. This comparison is performed for the ratios of $O(p^4)$ couplings.
Agreement with the fixed point values is found for a surprisingly large number of ratios, thus confirming
to a certain extent the fixed point solutions by data. 

A first reference to fixed points of the linearly realized chiral Lagrangian
was given in Ref.\,\cite{lin}. For a review on the role of fixed
manifolds in other contexts we refer to Ref\,\cite{schrem}. As fixed point solutions correspond to 
renormalization
group invariant relations between couplings, they may also be viewed as 
solutions of the parameter reduction program\,\cite{oehme}. Earlier 
applications of the
parameter reduction technique in the framework of effective field theories
may be found in\,\cite{atance}.

Let us first fix our notations for the couplings for $SU(3)$ and
$SU(2)$ symmetry, respectively. The effective chiral Lagrangian for chiral $SU(3)\times SU(3)$
symmetry, expanded in terms of increasing powers of $p^2$, may be written as follows
\begin{equation}
{\cal L}_\chi  =  {\cal L}^{(2)} + {\cal L}^{(4)} + {\cal L}^{(6)} + \cdots 
  =  {\cal L}^{(2)} + \sum_i L_i {\cal O}_4^i + \sum_i {K_i\over \Lambda^2} 
        {\cal O}_6^i + \cdots, 
\end{equation}
where ${\cal O}_4$ and ${\cal O}_6$ are dimension four and six
operators, respectively, and where the dimensionful cut-off $\Lambda$
is introduced, leading to dimensionless couplings $L_i$ and $K_i$. The 
couplings to external right-handed and left-handed vector fields $r_\mu, l_\mu$, and
scalar and pseudoscalar fields $s, p$ are included.
The lowest order, $O(p^2)$, Lagrangian ${\cal L}^{(2)}$ may be written 
as
\begin{equation}
{\cal L}^{(2)}_{SU(3)} = {F_0^2\over 4} \langle D_\mu U^\dagger D^\mu U\rangle +
                 {F_0^2\over 4} \langle U^\dagger\chi + \chi^\dagger U\rangle.
\end{equation}
The operation
$\langle\cdot\rangle$ denotes the trace, the external fields enter through 
$\chi = 2B_0(s+ip)$ and the covariant derivative   
$D_\mu U = \partial_\mu U -ir_\mu U + iUl_\mu$. The unitary matrix $U$
is given for $SU(3)\times SU(3)$ symmetry in terms of the Goldstone boson
fields $\Phi(x)$
\begin{equation}
U(\Phi) = \exp \left( i\sqrt{2} \Phi /F_0 \right)\ \ {\rm with}\ \ \Phi(x) \equiv \left( \begin{array}{ccc}
             {\pi^0\over \sqrt{2}}+{\eta\over\sqrt{6}} & \pi^+ & K^+ \\
             \pi^- & -{\pi^0\over \sqrt{2}}+{\eta\over\sqrt{6}} & K^0 \\
             K^- & \bar K^0 & -2{\eta\over\sqrt{6}}
                      \end{array} \right).
\end{equation}The two constants\,\cite{gass} $F_0, B_0$
are undetermined by the symmetry; they are related to the pion decay
constant and the quark condensate, respectively.

For the $O(p^4)$ Lagrangian ${\cal L}^{(4)}$ in the $SU(3)$ case we follow the conventions for
the couplings $L_i$ by Gasser and 
Leutwyler\,\cite{gass}
\begin{eqnarray}
{\cal L}^{(4)}_{SU(3)} & = & \sum_{i=1}^{10} L_i {\cal O}_i + 
\sum_{i=1}^2 H_i {\cal O}'_i \\
 & = & L_1 \langle D_\mu U^\dagger D^\mu U\rangle^2 + 
       L_2 \langle D_\mu U^\dagger D_\nu U\rangle
           \langle D^\mu U^\dagger D^\nu U\rangle + 
       L_3 \langle D_\mu U^\dagger D^\mu U D_\nu U^\dagger D^\nu U\rangle\nonumber \\ 
  & & + L_4 \langle D_\mu U^\dagger D^\mu U\rangle 
         \langle U^\dagger\chi + \chi^\dagger U\rangle 
      + L_5 \langle D_\mu U^\dagger D^\mu U
         (U^\dagger\chi + \chi^\dagger U)\rangle 
      + L_6 \langle U^\dagger\chi + \chi^\dagger U\rangle^2 \nonumber \\  
 & &  + L_7 \langle U^\dagger\chi - \chi^\dagger U\rangle^2  
      + L_8 \langle U^\dagger\chi U^\dagger\chi + 
                 \chi^\dagger U\chi^\dagger U\rangle  
     -iL_9 \langle F_R^{\mu\nu}D_\mu U D_\nu U^\dagger + 
         F_L^{\mu\nu}D_\mu U^\dagger D_\nu U\rangle \nonumber \\
& &   +L_{10} \langle U^\dagger F_R^{\mu\nu}UF_{L\mu\nu}\rangle 
     +\ {\rm terms\ involving\ external\ fields\ only},
\end{eqnarray}
where the field 
strenght tensors of the external gauge fields are
$F_{R\mu\nu} = \partial_\mu r_\nu - \partial_\nu r_\mu - i [r_\mu, r_\nu]$, 
and similarly for $F_{L\mu\nu}$. 

For the $O(p^4)$ Lagrangian ${\cal L}^{(4)}$ in case of $SU(2)\times
SU(2)$ symmetry we follow the conventions for
the couplings $l_i$ by Gasser and 
Leutwyler\,\cite{cou} 
\begin{eqnarray}
{\cal L}^{(4)}_{SU(2)} & = & l_1 (\nabla^\mu U^\top \nabla_\mu U)^2 +
   l_2 (\nabla^\mu U^\top \nabla^\nu U)(\nabla_\mu U^\top \nabla_\nu
   U)+ l_3 (\chi^\top U)^2 
+ l_4 (\nabla^\mu \chi^\top \nabla_\mu U)+\\
 & & + l_5 (U^\top F^{\mu\nu}F_{\mu\nu}U) + l_6 (\nabla^\mu U^\top F_{\mu\nu} \nabla^\nu U) 
   + l_7 (\tilde{\chi}^\top U)^2 + {\mathrm{terms\ involving\ external\ fields,}} \nonumber
\end{eqnarray}
where $U^A(x)$ is a four-component real $O(4)$ vector field of unit lenght,
$U^\top U=1$, with covariant derivative 
\begin{equation}
\nabla_\mu U^0 =  \partial_\mu U^0 + a^i_\mu(x) U^i,\ \ \ 
\nabla_\mu U^i  =  \partial_\mu U^i +\epsilon^{ikl}v^k_\mu(x)U^l
                     - a^i_\mu(x) U^0. 
\end{equation}
The external fields are $v_\mu(x)=v_\mu^i(x)\tau^i/2$,
$a_\mu(x)=a_\mu^i(x)\tau^i/2$, $s(x)=s^0(x)\boldmath{1}+s^i(x)\tau^i$
and $p(x)=p^0(x)\boldmath{1}+p^i(x)\tau^i$ and the vectors $\chi$ and
$\tilde{\chi}$ are given by $\chi^A = 2 B (s^0, p^i)$ and
$\tilde{\chi}^A = 2 B (p^0, -s^i)$.

The $O(p^6)$ couplings will be denoted by $K_i$ for $SU(3)$ and by
$k_i$ for $SU(2)$. There are too many to be spelt out here, they may
be found in Ref.\,\cite{nueva}.

Following the renormalization procedure for effective theories outlined
in the introduction, we use dimensional regularization and a mass
independent renormalization scheme. According to the renormalization program
for the chiral Lagrangian\,\cite{wei}, the
relations between bare and renormalized couplings are:
\begin{eqnarray}
SU(3):\ \ L_i^b & = & \mu^{(d-4)} \left[ L_i^r + \Gamma_i \lambda
\right],\label{1}\\
K_i^b & = & \mu^{2(d-4)} \left[ K_i^r + 
          \left( C_i + D_{ij} L_j^r \right) \lambda 
          + A_i\lambda^2 \right] ,\label{eq:doska}\\
SU(2):\ \ \ l_i^b &=&  \mu^{(d-4)} \left[ l_i^r + \gamma_i\lambda\right],\\
k_i^b & = & \mu^{2(d-4)} \left[ k_i^r + 
          \left( c_i + d_{ij} l_j^r \right) \lambda 
          + a_i\lambda^2 \right],
\end{eqnarray}
where $\lambda = [(d-4)^{-1}-\zeta]/16\pi^2$,
and $\zeta$ is a constant which depends on the renormalization
scheme. The constants $\Gamma_i,\ C_i,\ D_{ij},\ A_i$ for $SU(3)$ and
$\gamma_i,\ c_i,\ d_{ij},\
a_i$ for $SU(2)$ are calculable and known; they
are scheme independent. The constants $\Gamma_i$, $\gamma_i$ will be summarized
below, for the others we refer to Ref.\,\cite{nueva}. 

Unless a distinction between the $SU(2)$ and the $SU(3)$ cases is
necessary, in the following the capital letters $L_i,\ K_i,\
\Gamma_i,\ C_i,\ D_i$
are used as common fill-ins for the $SU(3)$ quantities and for the
$SU(2)$ quantities $l_i,\ k_i,\ \gamma_i,\ c_i,\ d_i $.
 
The renormalization 
group equations (RGE) for the renormalized couplings then have the
following generic form
\begin{eqnarray}
\mu{d L_i^r \over d\mu} & = & -{1\over 16\pi^2} \Gamma_i \label{Li},\\
\mu{d K_i^r \over d\mu} & = & -{1\over 16\pi^2} \left(C_i + D_{ij}L_j^r \right),\label{Ki}
\end{eqnarray}
where summation over repeated indices is implied. 
Henceforth we
shall drop the index $r$, since we shall exclusively deal with
renormalized quantities. 

Let us next proceed with an analysis of the
renormalization group equations (\ref{Li}) for the $O(p^4)$ couplings
$L_i$ for infrared fixed points
and their comparison to experiment. It is obvious that Eq.~(\ref{Li})
has no IR fixed point. However, the RGE for the ratios $L_i/L_j$ of
couplings 
\begin{equation}
\mu{d\over d\mu}\left({L_i\over L_j}\right)={1 \over 16\pi^2}{\Gamma_j \over L_j}
\left({L_i\over L_j} - {\Gamma_i\over \Gamma_j}\right) 
\end{equation}
have a non-trivial fixed point each at
\begin{equation}
\left. {L_i \over L_j} \right|_{{\mathrm f.p.}}  = {\Gamma_i \over
\Gamma_j}\ \ \ {\rm for}\ \ \ \Gamma_i,\ \Gamma_j\neq 0. \label{eq:ls}
\end{equation}
The general solution of the RGE (\ref{Li}) for initial value
$L_i(\Lambda)$ at $\mu=\Lambda$ is
\begin{equation}
L_i(\mu)=L_i(\Lambda)-{1\over 16\pi^2} \Gamma_i\;log{\mu\over
  \Lambda}.
\label{gensol1}
\end{equation}
Obviously, in the IR limit $\mu\rightarrow 0$ the ratios $L_i(\mu)/L_j(\mu)$ of {\it all general
solutions} approach the corresponding IR fixed
points (\ref{eq:ls}), controlled by the approach of $log(\mu/
\Lambda)\rightarrow -\infty$. 

It is now interesting to compare the experimental values for the
ratios of couplings with the predictions obtained 
for their IR fixed points (\ref{eq:ls}).

The following tables summarize the experimental values of the
couplings $l_i$ for $SU(2)$ symmetry\,\cite{gass,cou} at an energy
scale of the order of the pion mass and of the couplings $L_i$ for $SU(3)$
symmetry\,\cite{3cou} at an energy
scale $\mu = m_\rho$ (which is unfortunately rather large). These values are
derived from meson
decay constants, electromagnetic form factors and, for $SU(3)$ symmetry, 
also from semileptonic kaon decays. Also the corresponding  
coefficients $\gamma_i$ resp. $\Gamma_i$ of the beta functions are given.

\begin{center}
$SU(2)$:\ \ \begin{tabular}{|c|cccccc|} \hline
i & 1 & 2 & 3 & 4 & 5 & 6  \\ \hline
$l_i\cdot 10^3$ & $-2.4\pm 3.9$ & $12.7\pm 2.7$ & $-4.6\pm 3.8$ 
& $27.2\pm 5.7$ & $-7.3\pm 0.7$ & $-17.4\pm 1.2$ \\ 
$\gamma_i$ & 1/3 & 2/3 & -1/2 & 2 & -1/6 & -1/3 \\ \hline 
\end{tabular}
\end{center}
\begin{center}
$SU(3)$:\ \ \begin{tabular}{|c|ccccc|} \hline
i & 1 & 2 & 3 & 4 & 5 \\ \hline
$L_i\cdot 10^3$ & $0.4\pm 0.3$ & $1.35\pm 0.3$ & $-3.5\pm 1.1$ & 
$-0.3\pm 0.5$ & $1.4\pm 0.5$ \\ 
$\Gamma_i$ & 3/32 & 3/16 & 0 & 1/8 & 3/8\\ \hline
\end{tabular}
\end{center}
\hspace*{3.2cm}\begin{tabular}{|c|ccccc|} \hline
i & 6 & 7 & 8 & 9 & 10\\ \hline
$L_i\cdot 10^3$& $-0.2\pm 0.3$ & $-0.4\pm 0.2$ & $0.9\pm 0.3$ & 
$6.9\pm 0.7$ & $-5.5\pm 0.7$ \\ 
$\Gamma_i$& 11/144 & 0 & 5/48 & 
1/4 & -1/4 \\ \hline
\end{tabular}

\vspace*{5mm}

\begin{figure}[h]
\begin{center}
\leavevmode
\epsfxsize=12cm
\epsfbox{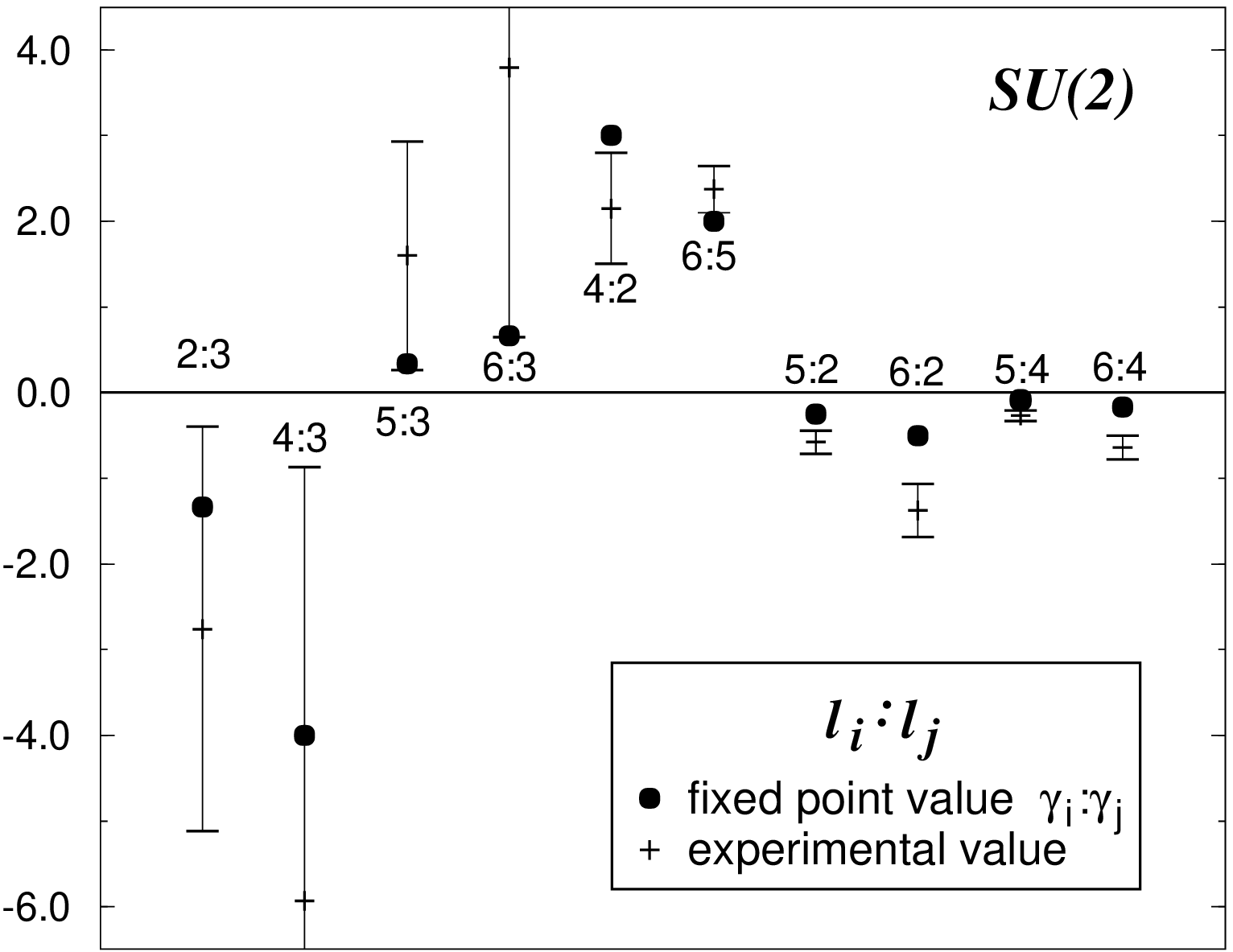}
\end{center}
\caption{Ratios $l_i/l_j$ of the experimental values of the couplings
$l_i$ in comparison with the corresponding IR fixed point ratios
$\gamma_i/\gamma_j$ for $SU(2)$ symmetry. To guide the eye, the ratios 
are ordered from left to right according to decreasing agreement.} \label{fig:su2}
\end{figure}

\begin{figure}[h]
\begin{center}
\leavevmode
\epsfxsize=14cm
\epsfbox{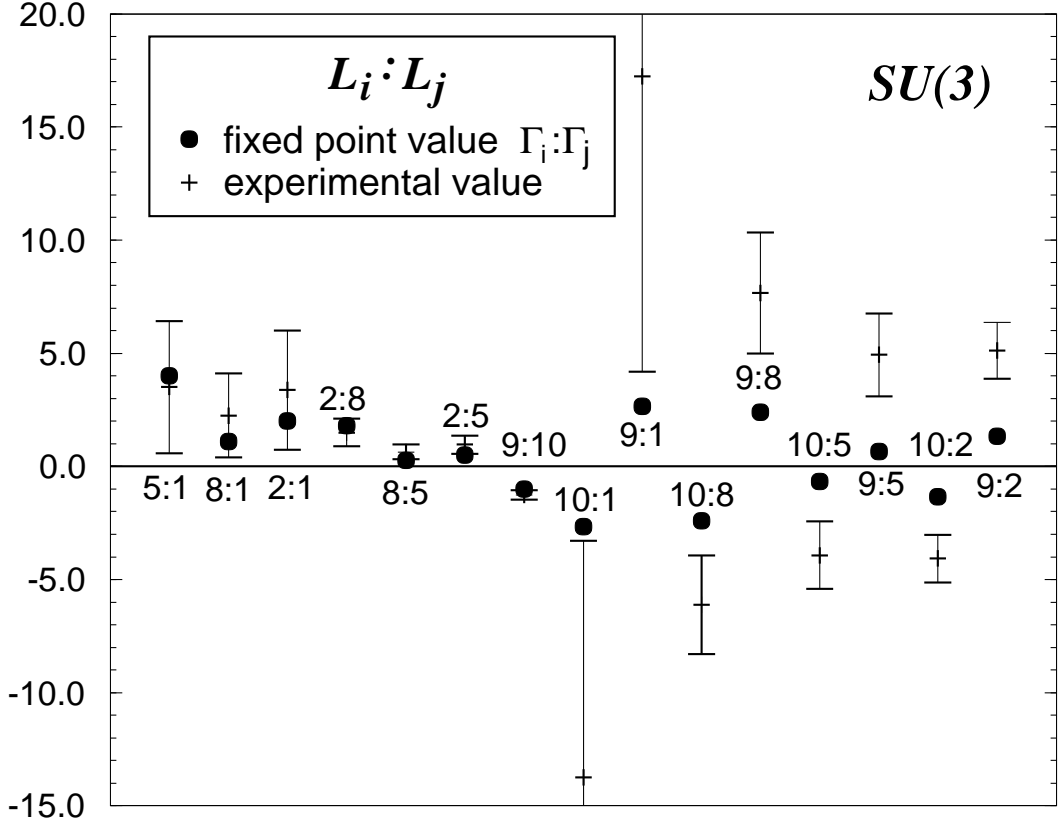}
\end{center}
\caption{Ratios $L_i/L_j$ of the experimental values of the couplings
$L_i$ in comparison with the corresponding IR fixed point ratios
$\Gamma_i/\Gamma_j$ for $SU(3)$ symmetry. To guide the eye, the ratios 
are ordered from left to right according to decreasing agreement.} \label{fig:su3}
\end{figure}

\smallskip

In Figs. \ref{fig:su2} and  \ref{fig:su3} the experimental ratios
$l_i/l_j$ for $SU(2)$ and $L_i/L_j$ for $SU(3)$ are compared  with
their respective fixed point ratios $\gamma_i/\gamma_j$ and
$\Gamma_i/\Gamma_j$ for values of $i,\ j$ for which $\gamma_i$,
$\gamma_j$, $\Gamma_i$ and
$\Gamma_j$ are nonzero. We omitted  $l_1$ for $SU(2)$  and 
$L_4$ and $L_6$ for $SU(3)$, since their errors are too large to give
meaningful comparisons. To guide the eye, we ordered the ratios from
left to right with decreasing agreement of experimental values with
the fixed point values. The overall agreement between experimental and
predicted values is quite impressive. Even where the agreement fails
on a quantitative level (at the right end of the figures), the sign and
the qualitative tendency are correct and the deviation is generically of
the order of one standard deviation.  

Encouraged by this amount of experimental support, we proceed to the analysis of the $O(p^6)$ RGE~(\ref{Ki}). 
The general solution of the generic RGE
(\ref{Ki}) for the $O(p^6)$ couplings $K_i(\mu)$ in terms of the general
solution for the $O(p^4)$ couplings $L_i(\mu)$ is 
\begin{equation}
K_i(\mu)=K_i(\Lambda)+\displaystyle{\frac{1}{2
    D_{ij}\Gamma_j}}\left(\left(C_i+D_{ij}L_j(\mu)\right)^2-\left(C_i+D_{ij}L_j(\Lambda)\right)^2\right).
\label{gensol2}
\end{equation}
The general solution as function of $\mu$ follows by inserting the
general solution (\ref{gensol1}) for $L_i$ into
Eq. (\ref{gensol2})
\begin{equation}
K_i(\mu)=K_i(\Lambda)-\displaystyle{\frac{1}{16\pi^2}}\left(C_i+D_{ij}L_j(\Lambda)\right)log\displaystyle{\frac{\mu}{\Lambda}}+\displaystyle{\frac{1}{2}}\displaystyle{\frac{1}{(16\pi^2)^2}}D_{ij}\Gamma_j\left(log\displaystyle{\frac{\mu}{\Lambda}}\right)^2.
\end{equation}
First of all we notice a renormalization group invariant relation
between the $O(p^6)$ couplings $K_i$ and the $O(p^4)$ couplings $L_j$,
{\it an IR attractive fixed line} in the plane of $K_i$ versus
$D_{ij}L_j$ 
\begin{equation}
K_i=\displaystyle{\frac{1}{2
    D_{ij}\Gamma_j}}\left(C_i+D_{ij}L_j\right)^2,\ \ {\rm valid\ for\
    all\ values\ of\ }\mu.
\label{fline}
\end{equation}
(or a fixed point in the variable $K_i/\left(C_i+D_{ij}L_j\right)^2$) 
or an infrared attractive fixed hypersurface in the space of $K_i$ versus the
contributing variables $L_j$.
This special solution of the RGE (\ref{Li},\,\ref{Ki}) is not
determined by initial value conditions; it is {\it IR attractive for
  all solutions} (\ref{gensol1}),(\ref{gensol2}), i.e. for the
renormalization group flow, in the limit $\mu\rightarrow 0$,
  i.e. $log(\mu/\Lambda)\rightarrow -\infty$. This is a very interesting result,
which predicts the $K_i$ in terms of the $L_i$. 

If one inserts
furthermore the fixed point solutions (\ref{eq:ls}), which implies
$D_{ij} L_j=D_{ij} \Gamma_j(L_m/\Gamma_m)$ for any suitable fixed value of $m$, into the fixed line
solution (\ref{fline}), one obtains by elimination of the coupling $L_m$ a renormalization group invariant
relation between any two $O(p^6)$ couplings $K_i$ and $K_j$
\begin{equation}
\sqrt{K_i}-\displaystyle{\frac{C_i}{\sqrt{2D_{ik}\Gamma_k}}}=\sqrt{\displaystyle{\frac{D_{ik}\Gamma_k}{D_{jk}\Gamma_k}}}\left(\sqrt{K_j}-\displaystyle{\frac{C_j}{\sqrt{2D_{jk}\Gamma_k}}}\right).\label{fline2}
\end{equation}
Again this fixed line in the plane $K_i$ versus $K_j$ is IR attractive for the RG flow. 
In the IR limit $\mu\rightarrow 0$, this fixed line approaches a constant
ratio for the ratios of the $O(p^6)$ couplings 
\begin{equation} 
\displaystyle{\frac{K_i}{K_j}}  \rightarrow \displaystyle{\frac{D_{ik}\Gamma_k}{
D_{jk}\Gamma_k}}\ \ \ {\rm for}\ \ \ \mu\rightarrow 0.\label{eq:kis}
\end{equation}

\begin{figure}[h]
\hspace*{-5mm}
\epsfig{file=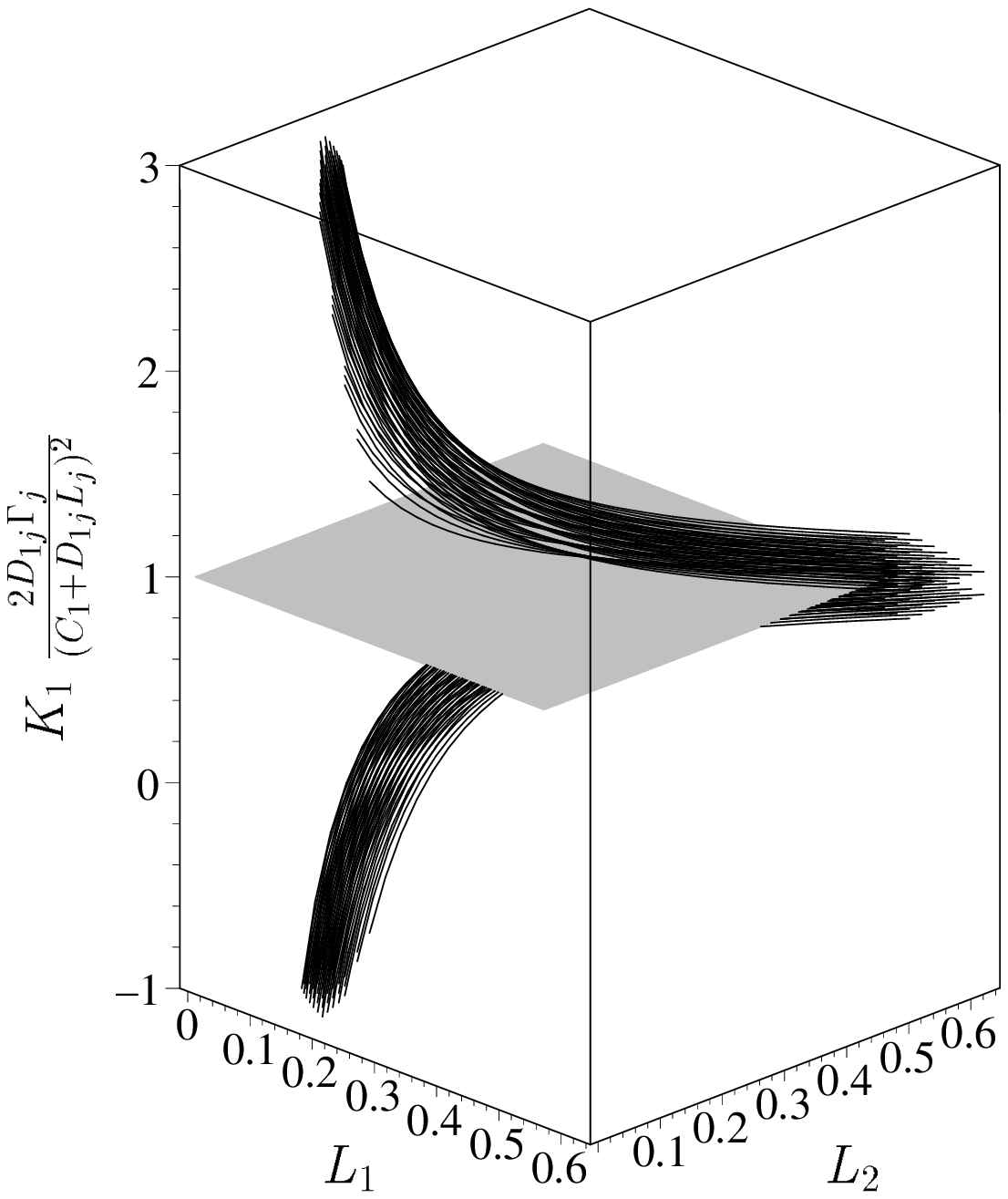,width=8.2cm,bbllx=96pt,bblly=203pt,bburx=436pt,bbury=577pt}\hspace*{8mm}\raisebox{-4mm}{\epsfig{file=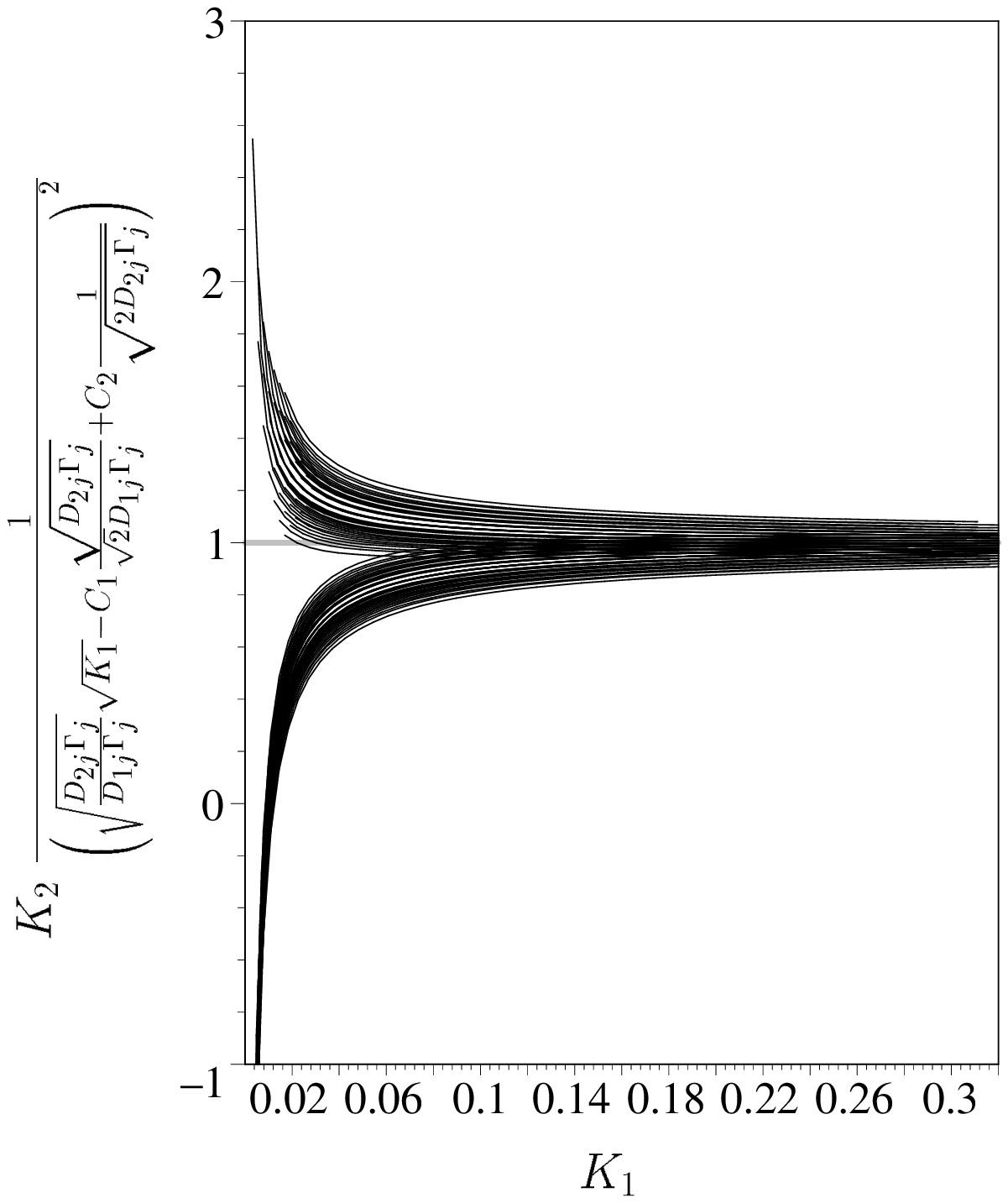,width=8.2cm,bbllx=79pt,bblly=172pt,bburx=436pt,bbury=577pt}}
\caption{Generic example, illustrating the IR attraction of the
  renormalization group flow towards the IR fixed manifolds which 
  relate $O(p^6)$ couplings
  to $O(p^4)$ couplings and $O(p^6)$ couplings among each other.}
\label{ill}
\end{figure}

The IR fixed manifolds (\ref{fline}) and (\ref{fline2}) are solutions
establishing renormalization group invariant relations between
couplings independently of any initial values. They may be similarly
relevant in nature as the IR fixed points (\ref{eq:ls}) for ratios
of $O(p^4)$ couplings which found substantial support by data. In addition the fixed manifolds (\ref{fline}) and
(\ref{fline2}) are IR attractive for the renormalization group flow in
its evolution from $\mu=\Lambda$ towards the IR. Even though this
evolution path is in practice not very long, let us illustrate in
Fig.\,\ref{ill} the
mathematical fact of this IR attraction (for a fictitious generic case
$\Gamma_{1,2}=1,\ C_{1,2}=1/(16\pi^2),\
D_{11}=D_{12}=D_{21}=1,\,D_{22}=2$ and suitable initial values). For
presentation reasons the
dependent $O(p^6)$ variables $K$ are normalized such that their IR
fixed value is equal to one. 

It has to be realized that if it were not for accidental zeros, the IR fixed
manifolds express all $O(p^4)$ and $O(p^6)$ in terms of a single
coupling. In practice, it will be a comparatively small number of
independent couplings.
 
The  relations (\ref{Li})-(\ref{eq:kis}) have been derived in the $SU(3)$ nomenclature with
capital letters. As pointed out earlier, they are valid also for
$SU(2)$ symmetry, if the capital letters are replaced by the small
ones. The relations hold for any of the relevant sets of couplings $L_i,\ K_j$
and $l_i,\ k_j$ with the numbers $\Gamma_i$ and $\gamma_i$ as given in
the tables and the numbers $C_i$, $D_{ij}$ and $c_i$, $d_{ij}$ as
given in Ref. \cite{nueva}. 
We found reasonable   
experimental support for the infrared fixed point solutions for ratios
of $O(p^4)$ couplings. There is, unfortunately, not sufficient experimental
information to subject the  infrared fixed line solutions between
$O(p^4)$ and $O(p^6)$
couplings and among $O(p^6)$ couplings to a similar test.\\[-3mm] 

{\bf Acknowledgements.} One of us (M.A.) would like to thank the Institut
f{\"u}r Theoretische Physik und Astrophysik der Universit{\"a}t zu Kiel for
hospitality, and the spanish MEC for a grant of the FPU programme. One 
of us (B.S.) is grateful to the theory group of DESY for continuous hospitality.

\end{document}